\documentclass[journal= wet]{CUP-JNL-DTM}%

\usepackage{graphicx}
\usepackage{multicol,multirow}
\usepackage{amsmath,amssymb,amsfonts}
\usepackage{mathrsfs}
\usepackage{amsthm}
\usepackage{rotating}
\usepackage{appendix}
\usepackage{ifpdf}
\usepackage[T1]{fontenc}
\usepackage{newtxtext}
\usepackage{newtxmath}
\usepackage{textcomp}
\usepackage{xcolor}
\usepackage{lipsum}
\usepackage[colorlinks,allcolors=blue]{hyperref}

\theoremstyle{definition}

\numberwithin{equation}{section}

\jname{Data/Math}
\articletype{RESEARCH ARTICLE}
\jyear{YEAR}

\begin{document}

\begin{Frontmatter}

\title[Posture Clip: Sit properly or I won't let you work]{Posture Clip: Sit properly or I won't let you work}

\author[1,2]{Arka Majhi}
\author[1]{Aparajita Mondal}

\address[1]{\orgdiv{Faculty of Information Technology and Communication Sciences}, \orgname{Tampere University}, \orgaddress{\city{Tampere}, \postcode{33100}, \state{Pirkanmaa},  \country{Finland}}}

\address[2]{\orgdiv{School of Forest Sciences}, \orgname{University of Eastern Finland}, \orgaddress{\city{Joensuu}, \postcode{80101}, \state{North Karelia},  \country{Finland}}}

\authormark{Majhi et al.}


\abstract{

Poor posture is a significant concern due to its detrimental effects on health and productivity. This paper presents a collar-clipped device called PostureClip, designed to restrict users from sitting and working at a bent angle, by blacking out the screen and resuming on correcting posture, thereby promoting better posture. The device integrates sensors and feedback mechanisms to provide real-time posture feedback to users.

To evaluate the effectiveness of PostureClip, a controlled experiment was conducted with participants (n=165) who were working on a laptop/PC for over 6 hours per day. The participants were randomly assigned to both the intervention group (IG1,n=54 ; IG2,n=55), which used the collar-clipped device, and the control group (CG, n=56), which did not use the device. IG1 didn't get feedback while IG2 got feedback from the device by notifying and further darkening the screen. The study was conducted in the office environment of the participants, for 4 weeks, and metrics such as posture angle, duration of bent angle, and user feedback were collected.

Analysis revealed significant improvements in posture angle (p<0.001) and significant reduction in bent angle duration (p<0.01) for participants' group using PostureClip with feedback and compared to the group without feedback and the control group (who were not intervened). The qualitative analysis of user feedback highlighted the device's ease of use, effectiveness in providing timely feedback, and positive impact on participants' awareness and habits regarding posture. These results indicate that PostureClip is an effective tool for promoting better posture during sedentary work.

}

\end{Frontmatter}

\section{Introduction}

In today's digital age, sedentary work has become a pervasive global phenomenon (\cite{Chandwani2019}). With the rise of technology and the shift towards knowledge-based economies, more people are spending their working hours seated at desks, engaging in computer-based tasks (\cite{Parry2013}). While this transition has undoubtedly yielded advantages in productivity and connectivity, it has also precipitated a notable health issue: the prevalence of poor posture (\cite{Vos2016}).

Poor posture, characterized by slouching, hunching over, and adopting unnatural spinal positions, has become alarmingly common among the global workforce (\cite{Vasavada2015}). Prolonged periods of sitting and inadequate ergonomic support can contribute to musculoskeletal disorders, discomfort, and reduced productivity (\cite{Edmonston2011, Maakip2016}). Moreover, the adverse effects of poor posture extend beyond the workplace, affecting individuals' overall well-being and quality of life. 

While poor posture is a global concern, it assumes a particularly worrisome dimension in India. India stands as one of the world's rapidly expanding economies, with a substantial portion of the workforce engaged in sedentary occupations (\cite{Maakip2016, Singh2020}). As the Indian economy transitions towards knowledge-based industries, the number of desk-based jobs is rising. This shift has prompted an increase in the prevalence of poor posture-related issues, impacting the health and productivity of millions of workers across the country (\cite{Chandralekha2022, Riskind1982}).

The implications of this problem are multifaceted. For individuals, poor posture can result in musculoskeletal discomfort, pain, and reduced work efficiency. Moreover, it can result in enduring health challenges, including chronic back pain and spinal disorders(\cite{Reeve2009, Bener2004}). From an organizational perspective, poor posture contributes to decreased employee morale, increased absenteeism, and higher healthcare costs (\cite{Wynne2014, Patterson2018}). Therefore, tackling the problem of poor posture in the workforce is imperative not only for the health and welfare of individuals but also for the economic advancement and competitiveness of the nation.

Various interventions and technologies have been developed to combat the detrimental effects of poor posture. Among these, wearable devices that provide real-time feedback and promote posture correction have shown promise (\cite{Matuska2020, King2013, Zeemp2016, Yoong2019}). However, many existing solutions suffer from limitations such as discomfort, lack of long-term adherence, and a lack of specificity for sedentary work environments (\cite{Palsson2019, Harvey2020}). Therefore, there is a need for innovation and a research gap to find context-specific approaches that can effectively address the problem of poor posture. In response to this gap, this research aims to design and develop an innovative collar-clipped device called PostureClip. PostureClip aims to provide real-time feedback and encourage proper posture alignment during sedentary work activities. By leveraging sensor technology, user-centered design principles, and ergonomic considerations, PostureClip seeks to address the challenges associated with maintaining good posture in the workplace.

\subsection{Research Question and Contribution} 
We hypothesized that PostureClip could significantly change the posture of desk workers working on laptops or PCs through nudging and habit formation. We wanted to understand the impacts and subsequent behavioral changes on managing posture. We tested PostureClip with n=54 participants. We seek to learn through this study, after using PostureClip for a month at the workplace: (RQ1) Is there is a significant change in the participants' posture? and (RQ2) What are the differential effectiveness of changes across various participant categories? 

This research contributes to the theme of supporting worker well-being and health among several themes of CHI Work. It is a quasi-experiment-based study on how technologies can support maintaining the right posture required for desk-based jobs. The major contributions of this study to the CHI-Work and HCI community are: 1) Understanding the behavior of desk-based employees towards their posture, while working with laptop/Pc or screens 2) How a designed device like PostureClip can be effective in helping common postural issues prevalent. PostureClip addresses common postural issues, reducing specific posture deviations, such as forward head posture. 

This paper presents the comprehensive design, development, and evaluation of the PostureClip device. The subsequent sections will delve into the background of the problem, explore related works in the field of posture correction devices and wearable technologies, discuss the significance of ergonomic interventions, and provide a detailed account of the design and evaluation processes. Additionally, this paper will present the results of user studies conducted with desk-based workers, shedding light on the device's effectiveness, user satisfaction, and potential for mitigating the problem of poor posture in the workforce.

\section{Background and Related Works}
\subsection{Prevalence and Severity of Poor Posture} 

Damage to bodily structures, including muscles, joints, ligaments, tendons, nerves, bones, and localized blood circulation systems, is largely influenced or worsened by work and environmental conditions, categorizing them as work-related musculoskeletal disorders (WMSDs) (\cite{Podniece2008}). Studies conducted in Europe revealed that about three out of five employees experienced discomfort linked to WMSDs, highlighting their prevalence as the most widespread work-related health problem in the European Union (\cite{Podniece2008, EUWorkingCondition2016}). Backache is the predominant WMSD, affecting 43\% of all workers (\cite{Podniece2008}). Symptoms associated with WMSDs, such as musculoskeletal pain and discomfort, are prevalent among Visual Display Terminal (VDT) users, ranging from 7 to 30\% (\cite{Swinton2017}). These symptoms represent a global occupational health concern (\cite{Dianat2016, Maakip2016}). Occupational exposures significantly contribute to the development of WMSDs, with up to 37\% of all back pain and injuries linked to these factors, although this prevalence varies among countries (\cite{Punnett2005}). A study conducted within the university community in Nigeria identified that 75.7\% of respondents experienced shoulder and upper limb musculoskeletal symptoms due to poor posture and gadget use (\cite{Susilowati2022}). Furthermore, improper posture can result in musculoskeletal discomfort across different body regions, such as the neck, shoulders, and lower back, impacting the spine, muscles, tendons, joints, and bones (\cite{kim2015}). Maintaining postures that stress the body, such as prolonged slouching, heightens spinal pressure, rendering it more susceptible to injury and degeneration (\cite{Physiopedia2024}). 

In developing countries, poor posture is amplified by rapid economic expansion and proliferation of desk-oriented professions. This is likely to increase in countries transitioning from labor-intensive sectors to knowledge-based industries like information technology, finance, and business process outsourcing (\cite{Singh2013, Bosworth2007}). A study conducted among software professionals revealed that 67\% of participants reported experiencing musculoskeletal issues (\cite{Shrivastava2012, Singh2020}). Among these professionals, 59\% reported body aches, with backaches afflicting 34\% of individuals (\cite{Prasad2020}). These discomforts predominantly stemmed from prolonged sitting, inadequate posture, and sub-optimal ergonomic practices.

\subsection{Existing Posture Correction Devices and Wearable Technologies}  

Researches on posture correction devices and wearable technologies tried to address the issue of poor posture. Ailneni (\cite{Ailneni2019}) employed a sensor-based posture-correction wearable that detects changes in sitting posture and provides vibrotactile feedback to the user. While it might be effective in producing short-term improvements, the approach was evaluated with a limited sample size over a constrained period, and raised concerns regarding comfort, calibration, and sustained user compliance. Otoda (\cite{Otoda2018}) introduced the Census sensory chair, capable of classifying 18 sitting positions using data from eight accelerometers. However, this system’s dependence on multiple calibrated sensors increases cost, complexity, and sensitivity to placement variations, potentially limiting scalability across diverse seating contexts. Zemp (\cite{Zeemp2016}) augmented a standard office chair with multiple embedded sensors for motion detection. However the design requires frequent recalibration and can yield inconsistent readings across users and seating conditions. Huang (\cite{Huang2017}) developed a 0.255 mm-thick piezoresistive sensor matrix comprising two layers of polyester film for noninvasive posture monitoring. Although the thin-film design enables unobtrusive integration, it is prone to drift, hysteresis, and limited spatial resolution, which can hinder fine-grained posture discrimination in real-world settings.

However, despite their potential, many existing posture correction devices have limitations that hinder their widespread adoption and long-term efficacy. Issues such as discomfort, inconvenience, lack of real-time feedback, and low user adherence have been reported. Furthermore, these devices often lack customization for specific work environments, such as desk-based work, where individuals face unique challenges in maintaining good posture (\cite{Simpson2019}).

Following the review of existing studies and devices addressing posture-related issues, our research introduces a novel approach that effectively promotes proper sitting posture in a convenient and user-friendly manner. Our proposed device distinguishes itself by its efficiency, affordability, comfort, and active feedback mechanism. Unlike passive devices that rely on user interaction to address poor posture, our device provides real-time feedback and actively intervenes by automatically deactivating electronic devices when detecting slouching. Additionally, its compact and discreet design allows for seamless integration into the user's work life while seated at their desk. It serves as a reminder to maintain optimal posture throughout extended periods of sedentary activity. This active intervention strategy aims to instill positive posture habits effortlessly, contributing to long-term postural health and well-being.

\subsection{Habit formation theories}

The Operant Conditioning theory serves as a foundational framework for our study, offering valuable insights into the process of habit formation and behavior modification. This theory, pioneered by B.F. Skinner delineates a systematic approach to shaping behaviors by strategically applying reinforcements and punishments. By understanding how individuals respond to external stimuli and consequences, we can effectively manipulate environmental factors to promote desired behaviors and discourage undesirable ones (\cite{Tangmanee2021, Donahoe1988}).

In the context of our research, the Operant Conditioning theory is particularly relevant in addressing the prevalent issue of poor posture among computer users. By leveraging principles of reinforcement and punishment, we aim to cultivate a habit of maintaining proper posture during prolonged computer use. By consistently implementing reinforcements, such as rewards for upright posture and restrictions for slouching, we seek to induce a lasting behavioral change conducive to musculoskeletal health.

\section{Method}

\begin{figure}
\centering
\begin{tabular}{ccc}

\includegraphics[width=0.45\linewidth, angle =-90]{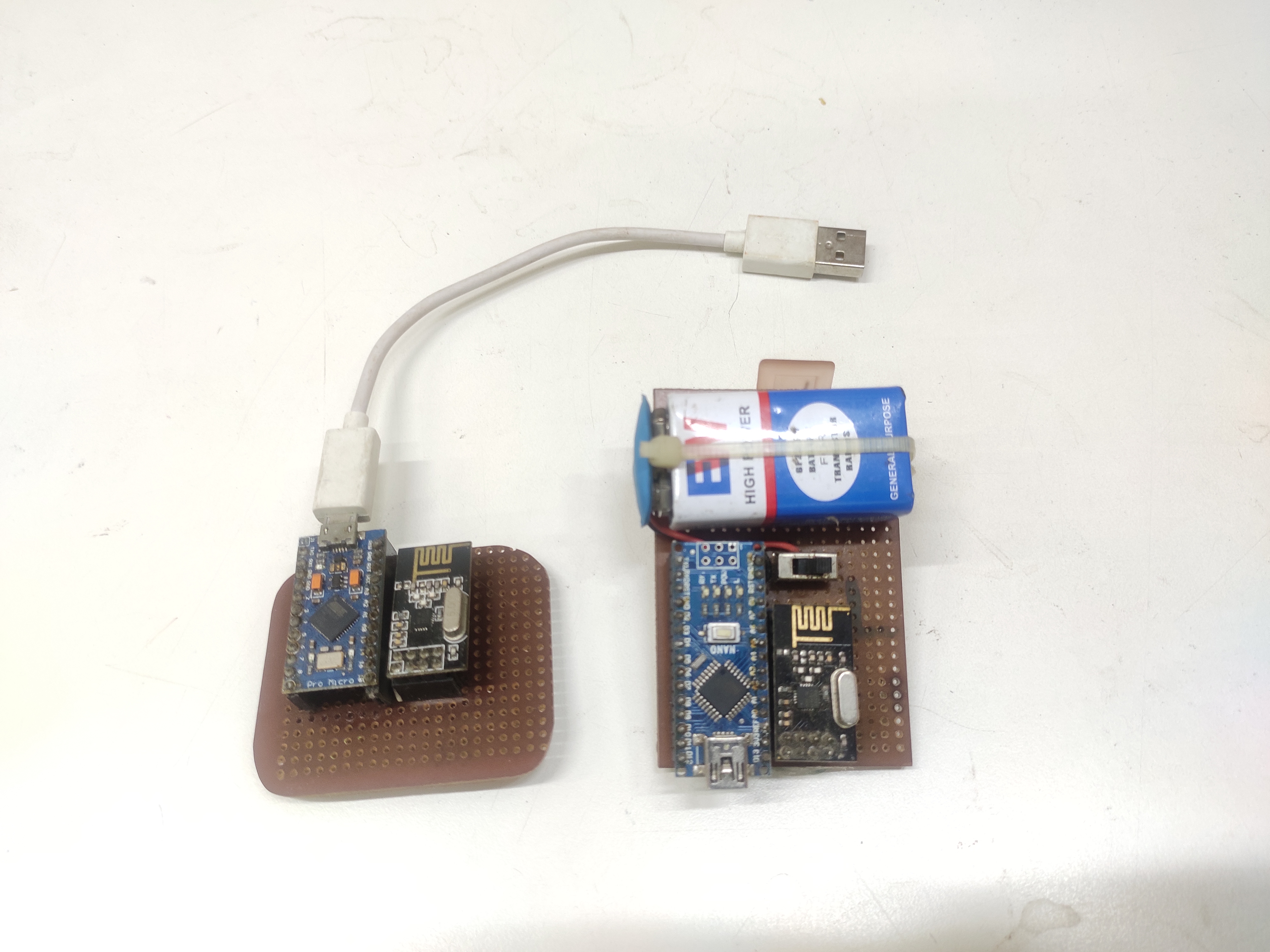}
&
\includegraphics[width=0.45\linewidth, angle =-90]{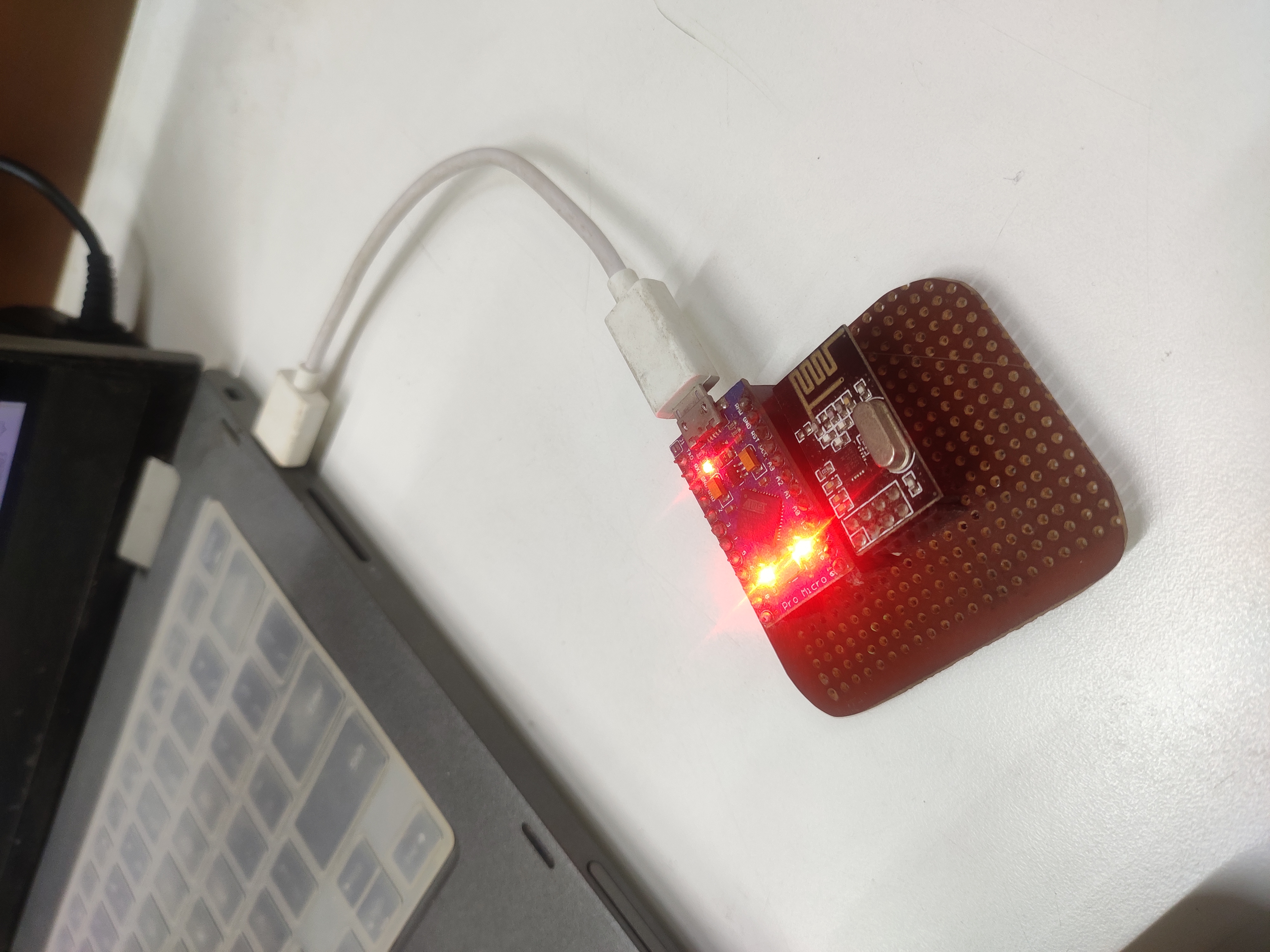}
&
\includegraphics[width=0.45\linewidth, angle =-90]{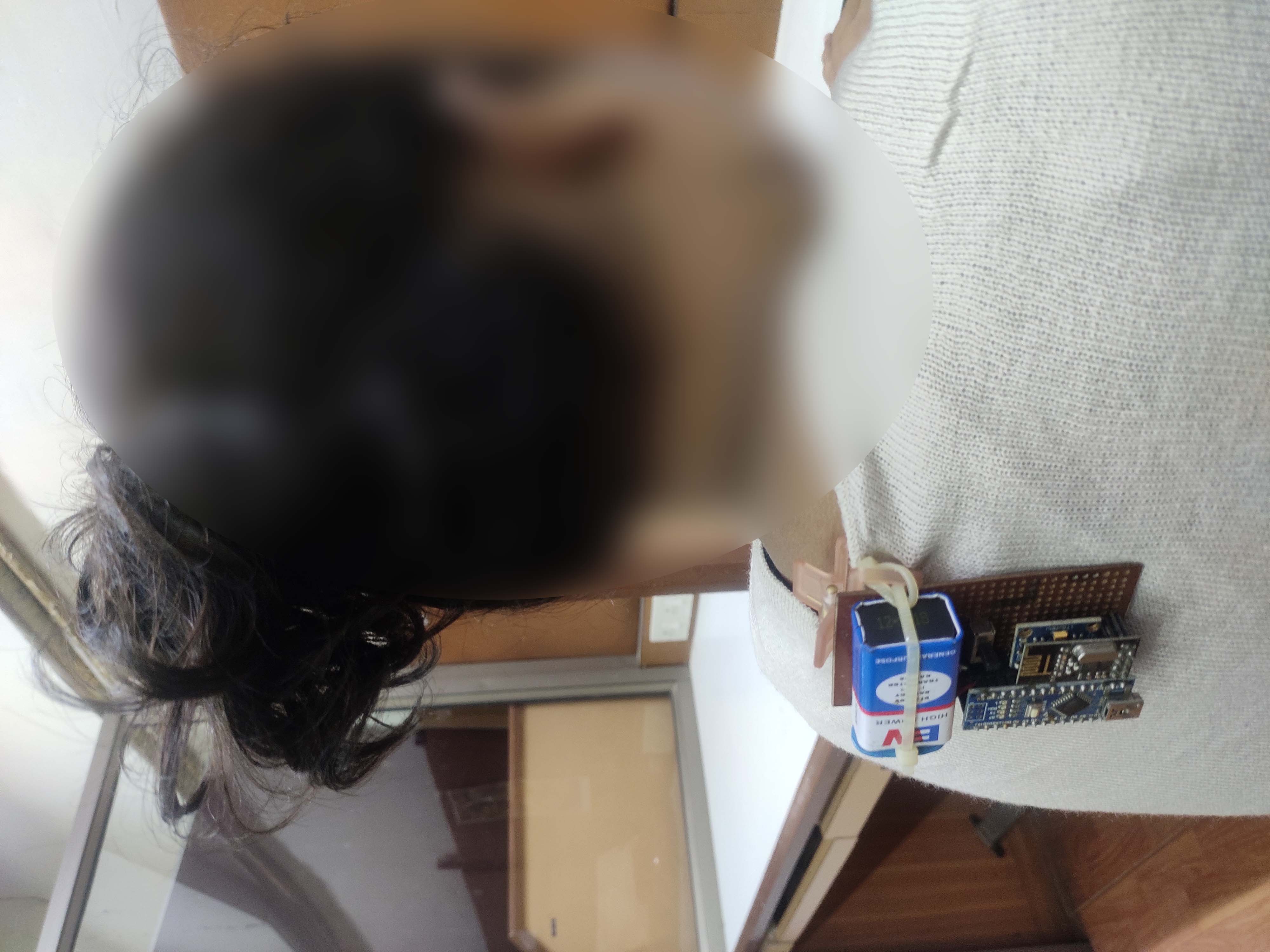}

\end{tabular}

\caption{Left: Receiver and Transmitter Module, Middle: Receiver module connected to a laptop, Right: Transmitter module clipped to collar or Participant}
\label{fig:receivertransmitter}
\end{figure}

\begin{figure}
    \centering
    \includegraphics[width=1.05\linewidth]{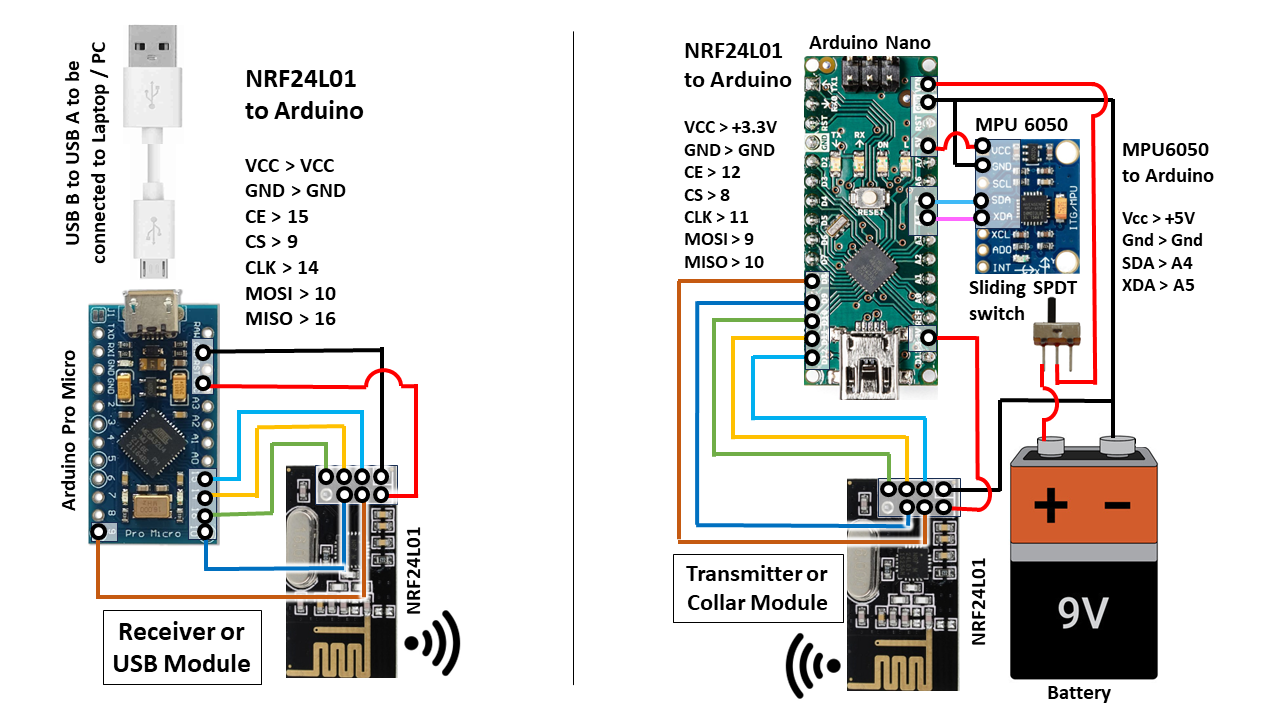}
    \caption{Left: Schematics of Receiver or USB Module ; Right: Schematics of Transmitter or collar clip module}
    \label{fig:receiver_transmitter_schematics}
\end{figure}

\subsubsection{Device Functionality}
The PostureClip device actively monitors and provides real-time feedback on the user's posture. By continuously measuring the neck and head tilt, it can detect deviations from optimal posture angles and provide corrective notifications to the user.

\begin{figure}
\centering
\begin{tabular}{cc}

\includegraphics[width=0.5\linewidth, angle =0]{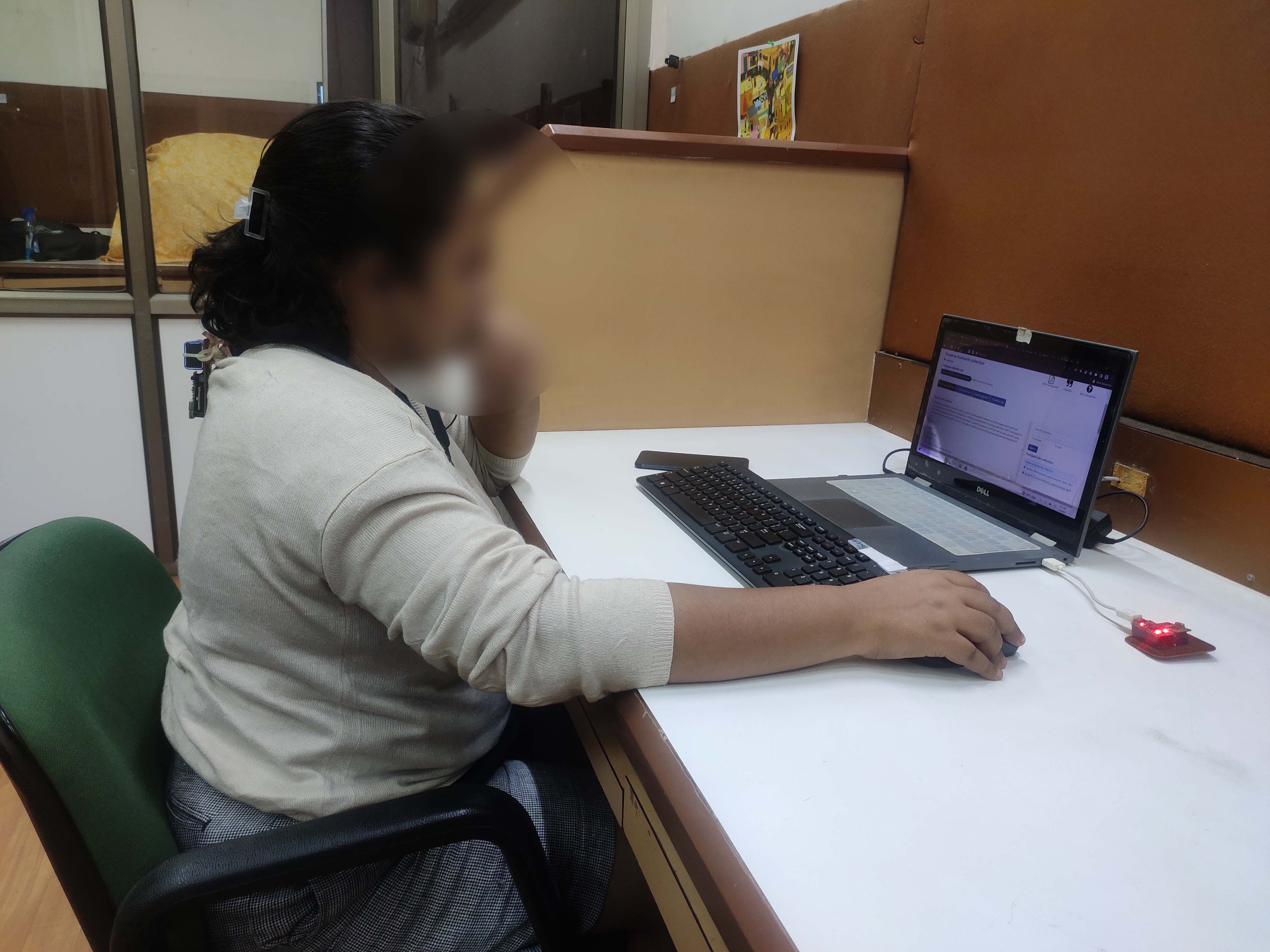}
&
\includegraphics[width=0.5\linewidth, angle =0]{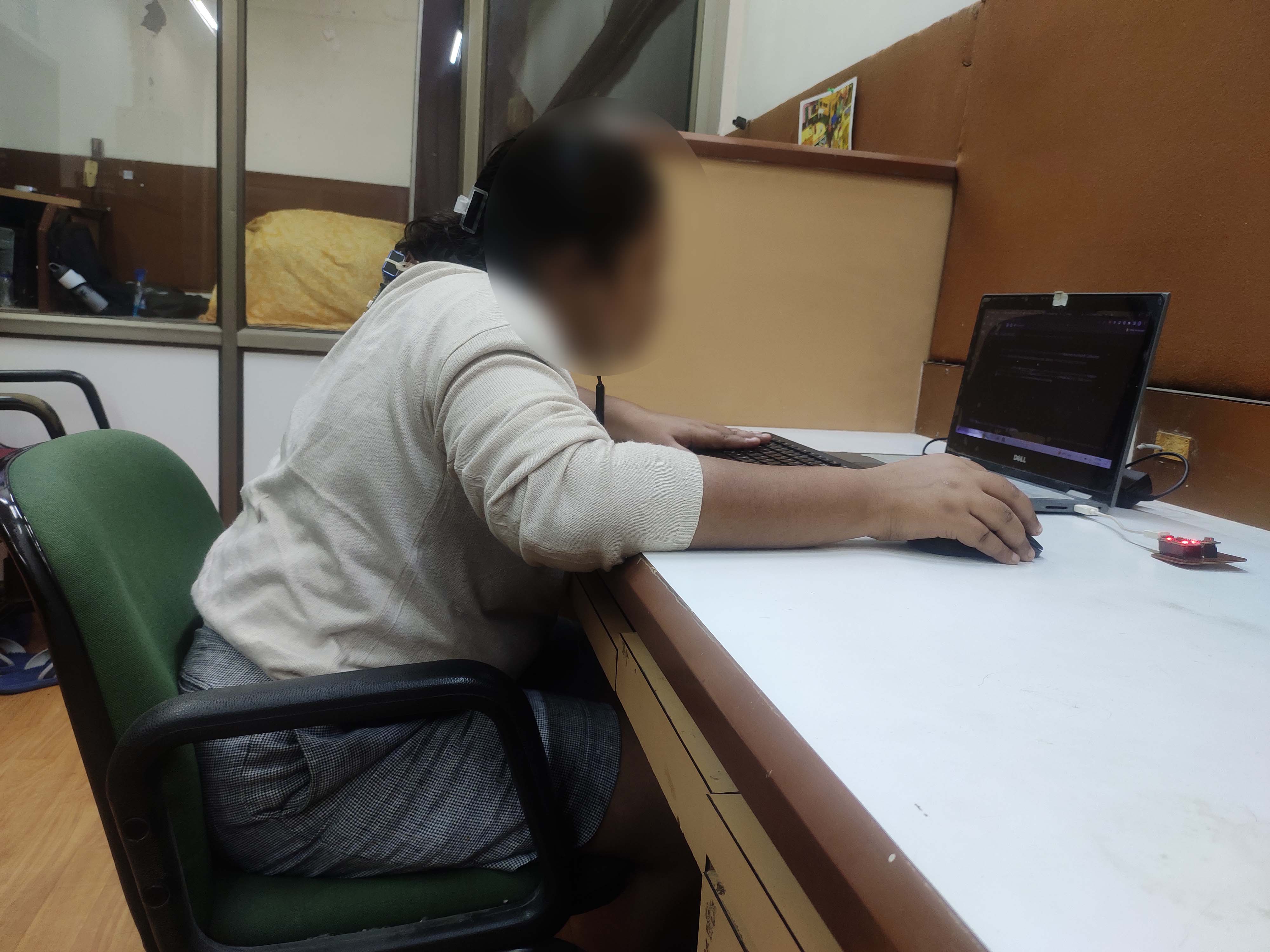}

\end{tabular}

\caption{Left: Female participant sitting straight and working on her laptop, Right: The same participant sitting bent for a minute, and the display turns off, nudging her to sit straight to continue work}
\label{fig:femaleparticipant}
\end{figure}

The device employs an algorithm that analyzes the sensor data to determine the user's posture alignment. This algorithm considers predefined posture thresholds and angles, comparing them with the real-time measurements obtained from the accelerometer and gyro meter. When a deviation or tilt is detected, the device triggers a gentle vibration to remind the user to adjust their posture. If the user continues working and does not correct the posture for over a minute, the display of the laptop or monitor turns dark until the user sits properly. This activity is demonstrated through images in Figure \ref{fig:femaleparticipant}.

Furthermore, the PostureClip device records posture data over time, allowing for a comprehensive analysis of posture patterns and trends. The data is analyzed, providing valuable insights into participants' posture habits.

\subsection{Device Development Process}
The development of the PostureClip device involved several stages, including hardware design, software programming, and calibration. The following steps outline the development process:

\subsubsection{Hardware Design}
The device's hardware components were selected based on their suitability for posture monitoring and user comfort. The Arduino microprocessor, MPU 6050 sensor, NRF24L01 module, and power source were carefully integrated into a compact, lightweight design that could easily be clipped onto a shirt collar towards the back. The device has two modules: the transmitter and the receiver. The transmitter collects the tilt angle in XYZ axes, processes the information, and transmits the raw data into data packets to the receiver. The receiver collects the data packets, and the algorithm processes the data. We developed 200 such devices for testing with the participants.

The current design prototype weighs 60 grams, including the battery. Implementing next-generation Arduino boards, like ESP8266/32 powered by a LiPo battery or integrated M5 stack modules, would reduce the device's weight as they have embedded sensors. Due to unavailability and resource constraints, we could not implement this hardware in the current study. Later studies can explore this possibility, which might reduce the discomfort of usage and the feeling of self-consciousness about its visibility in professional settings.

\subsubsection{Software Programming}
The code or algorithm for the device was developed using the Arduino programming language. The code was designed to read and process the sensor data, implement posture analysis algorithms, and enable wireless communication with the central monitoring system. Depending on the tilt angles received by the receiver, the algorithm calculates if the angle crossed the threshold angle set by the user during setup. Depending on the result, it either notifies and prompts the user to sit properly or stay idle. If the user ignores the notification for over a set time (typically over two minutes), the screen slowly turns dark until the user sits properly to resume working. The darkening of the screen and notification alerts are paused if the sensor detects jerks or movements or specific work conditions like walking, having tea, talking to colleagues, collaborating, drawing, note-taking, getting up, etc. The feedback interval (nudges every two minutes of sustained poor posture) was determined based on existing studies demonstrating the efficacy of frequent, timely prompts for breaking sustained unhealthy behaviors. 

\subsubsection{Calibration and design iterations}
Before conducting a session, the accuracy and reliability of the device for measuring tilt angles were calibrated with the virtual spirit level tool in the compass app on three Android phones for cross-checking. The measurements were validated daily by one of the authors, an expert in this field, and trained by a rehab specialist. User trials were conducted to gather feedback on comfort, the effectiveness of feedback mechanisms, and overall user experience. Based on the test results and user feedback, necessary iterations were made to enhance the device's performance and usability.

Integration with Monitoring System: The PostureClip device was seamlessly integrated with a central monitoring system, allowing for real-time data transmission and comprehensive analysis. The monitoring system was developed to receive and process the posture data, visualize the results, and provide insights to the users and researchers.

\subsection{Participants}

The sample size required for the study was calculated using G* power 3.0 software (\cite{Erdfelder2007}). In this study, 't-test' was chosen as the test family, 'Means: Difference between two different means (matched pairs)' was chosen as the statistical test for comparing pre and post-scores of IG, 'Means: Difference between two independent means (two groups)' was chosen as the statistical test for comparing two means (CG and IG) and (IG1 and IG2) and the medium effect size was d = 0.5. According to the calculation, a sample of around 55 participants in each group would be required in each IG1,IG2 and CG to achieve 99\% power to detect a difference at 0.05 significance level. Considering the attrition rate of 10\%, the calculated sample size required to have a significant effect would be 182 participants.

A total of 187 participants were recruited for the study using convenience sampling. Out of them, 165 were retained till the end of the study. The data from those 22 participants who left were excluded from quantitative analysis, but some of their interview reports were included in qualitative thematic analysis. The statistical details of participants are mean (SD) age, height, and body weight were 30.0 (8.5) years, 165.2 (9.4) cm, and 76.3 (13.1) kg, respectively. The inclusion criteria included employees aged 25-45 years working on laptops or PCs for at least 6 hours daily. Participants from various industries, including information technology, finance, and healthcare, were selected to ensure diverse sample representation. Due to COVID-19 norms, (n=26) participants also worked from home with a desk setup similar to that in their office. 

None of the participants has reported a previous history of surgery in the hips, backbone, shoulder, or neck. Informed written consent by signing was obtained from all participants before conducting the study. Participants who consented by signing a waiver form were photographed, and the images were stored as records. Some photographs of participants are used in this research publication with their due permission. Exclusive permission was taken from the employers of the IG2 group as the experiment involved turning off the screens or darkening them and preventing them from working until the participant sat in the right posture to resume working.

Of the 109 participants in IG, 54 (30 male, 24 female) were randomly selected for IG1, 55 (32 male, 23 female) were selected for IG2, and 56 (32 male, 24 female) were selected for CG. In all three groups, we tried to achieve a balanced distribution and representation of gender, age, the industry in which they worked, and their places of work (office or work from home). The baseline test shows no significant pre-existing differences between the three groups (p>0.05). No outliers were found.

The IG1 and IG2 groups were intervened with PostureClip, but only IG2 had the feedback mechanism for darkening the screen and not letting the user work until the posture is corrected. IG1 had been intervened with PostureClip without feedback to check for a placebo effect of the device or the intervention. CG was controlled, not given any devices, and checked for changes in posture behaviour after four weeks. The choice of a four-week study period was guided by practical feasibility and workplace constraints. Securing extended access to participants' work environments for longer durations was not always possible due to organizational policies and operational demands. The four-week period allowed for the capture of early-stage behavioral changes and preliminary habit formation, as suggested by habit formation theories that indicate noticeable behavioral shifts can emerge within a few weeks of consistent reinforcement.

\subsection{Data Collection Procedures}
Data collection involved two main components: objective posture measurements and subjective assessments through user feedback.

\subsubsection{Objective Posture Measurements}
This is a quasi-experimental study. Participants wore the PostureClip device on their workdays, five days a week, excluding holidays, for a span of four weeks. The device continuously collected tilt angle measurements from 11 AM to 4 PM at the collar point. The participants turned them on for about Mean(SD)-5 hours 25 minutes (42 minutes). These measurements were recorded continuously and synchronized with the online server for storage and analysis. A sample session is shown in Figure \ref{fig:femaleparticipant}.

Along with these other indicators, neck flexion, shoulder elevation, and lumbar curvature were measured by orthopedic doctors and professional physical therapists. Occurrences of forward head posture and rounded shoulders were also calculated. These measurements were conducted at the baseline and at the endline of the intervention for all groups.

\subsubsection{Subjective User Assessments}
After four weeks, semi-structured interviews were conducted with a subset of participants (CG,n=22; IG1,n=25; IG2,n=26) to gather qualitative insights into their experiences, challenges, and perceptions regarding the device's usability and effectiveness. User satisfaction was assessed using a post-intervention questionnaire that included the ten-item System Usability Scale (SUS) as well as structured Likert-scale (1–5) items covering perceived comfort, device interference, feedback utility, and likelihood of continued use. These standardized metrics were supplemented with open-ended questions for qualitative insight, consistent with best practices for evaluating wearable health technologies. Key insights included a high average SUS score (73/100, indicating good usability) and positive ratings for comfort and feedback relevance. However, some users (n=6) cited annoyance and distraction during frequent interventions, particularly during visually intensive work. Such findings align with prior studies indicating usability and adherence challenges in the deployment of wearable health technology. The interviews followed a semi-structured guide that explored user satisfaction, perceived benefits, barriers to adherence, and suggestions for improvement of user experience regarding usage of the device. We stopped interviewing and stopped at the numbers when we realized that the information or insights were getting saturated.

\section{Analysis and Findings}

\subsection{Quantitative Analysis}
The quantitative analysis aimed to examine the effectiveness of the PostureClip device in promoting proper posture alignment among participants. The analysis focused on changes in posture angles and the frequency of specific posture deviations before and after using the device.

\subsubsection{Posture Angles}

The data collected were checked for normality or normal distribution. Kolmogorov–Smirnov test was chosen as the sample size (IG) was greater than 50 (IG1,n=54 ; IG2,n=55 ; CG,n=56) (\cite{Mishra2019}). p-value was greater than 0.05 for all the factors, signifying normal distribution and satisfying conditions for parametric tests. Paired t-tests were conducted to determine the significance of the changes in posture angles for both IG1 and IG2. The results showed a statistically significant decrease in neck flexion in IG2 (t= -4.22, p < 0.001), indicating improved neck alignment. Additionally, in IG2, shoulder elevation was significantly reduced (t = -2.93, p = 0.005) and improved lumbar curvature (t = 3.11, p = 0.003) after using the device with feedback of turning off the display. In IG1 and CG, the mean differences were not statistically significant.

\begin{table}[htbp]

\centering
\caption{Posture-Related Measurements Across Groups: Pre-Post Intervention}
\label{tab:all_groups_results}
\scalebox{0.9}{
\begin{tabular}{l c c c c l}
\toprule
\textbf{Measure} & \textbf{Group} & \textbf{Before} & \textbf{After} & \textbf{Change} & \textbf{Interpretation} \\
\midrule
Neck Flexion Angle (\textdegree) & CG & 50.0 & 49.5 & -0.5 & No significant improvement \\
 & IG1 & 50.1 & 49.8 & -0.3 & Slight change without active feedback \\
 & IG2 & 50.3 & 45.0 & -5.3 & Significant improvement with feedback \\
\addlinespace
Shoulder Elevation (\textdegree) & CG & 30.0 & 29.8 & -0.2 & No meaningful change \\
 & IG1 & 30.5 & 30.2 & -0.3 & Minimal change without feedback \\
 & IG2 & 30.2 & 27.0 & -3.2 & Significant reduction with feedback \\
\addlinespace
Lumbar Curvature (\textdegree) & CG & 40.0 & 40.5 & +0.5 & Stable posture with no significant change \\
 & IG1 & 39.8 & 40.0 & +0.2 & No substantial change without feedback \\
 & IG2 & 39.9 & 43.0 & +3.1 & Noticeable improvement with feedback \\
\addlinespace
Forward Head Posture (\%) & IG1 & 45 & 43 & -2 & Slight reduction without feedback \\
 & IG2 & 47 & 20 & -27 & Significant decrease with feedback \\
\addlinespace
Rounded Shoulders (\%) & IG1 & 50 & 48 & -2 & Minor decrease without feedback \\
 & IG2 & 52 & 25 & -27 & Substantial reduction with feedback \\
\bottomrule
\end{tabular}}
\end{table}

\begin{figure}
\centering
\includegraphics[width=0.95\linewidth, angle =0]{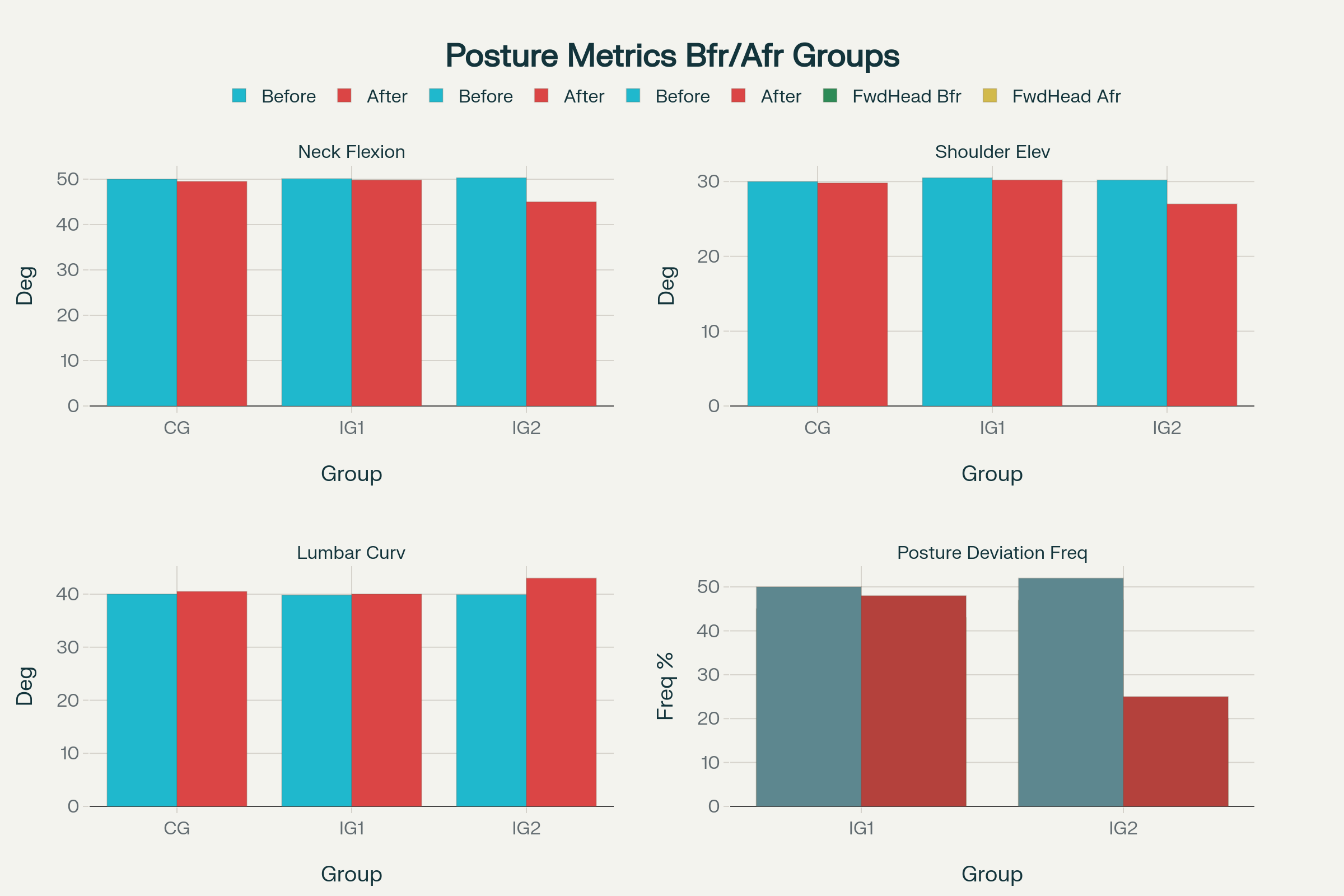}

\caption{Bar charts showing posture angles before and after intervention for groups CG, IG1, IG2 and posture deviation frequencies for IG1 and IG2 groups}
\label{fig:chart}
\end{figure}

\subsubsection{Frequency of Posture Deviations}
Multiple group Chi-square tests were conducted to examine the frequency of specific posture deviations before and after using the PostureClip device for the IG1 and IG2. The results revealed a significant reduction in the occurrence of forward head posture (X²(1) = 10.24, p = 0.001) and rounded shoulders (X²(1) = 6.78, p = 0.009) after using the device. This indicated that the device with feedback had a positive effect in reducing these common posture deviations.

\subsubsection{Correlation Analysis}
The correlation analysis explored the relationship between adherence to wearing the PostureClip device and changes in posture. The results showed a significant positive Pearson's correlation (r²=0.78) between the duration of wearing the device (measured in hours per day) and change in posture angles. IG2 Participants (n=55) who wore the device more consistently than IG1(n=53) during office hours experienced positive changes in neck flexion (r = 0.45, p = 0.001), shoulder elevation (r = 0.39, p = 0.006), and lumbar curvature (r = 0.48, p < 0.001).

\subsection{Qualitative Analysis}
The qualitative analysis aimed to gain deeper insights into participants' experiences, challenges, and perceptions regarding the usability and effectiveness of the PostureClip device. Thematic analysis was conducted to identify key themes and patterns in the qualitative data following the method of Boyatzis (\cite{Boyatzis1998}).

\subsubsection{User Satisfaction}
One of the prominent themes from the interviews was user satisfaction with the PostureClip device. Participants expressed overall satisfaction with the device, highlighting its effectiveness in increasing their posture awareness and promoting conscious adjustments throughout the day. While frequent interventions may maximize corrective feedback, several participants reported annoyance or screen blackout fatigue, especially during tasks that require intentional posture deviations (e.g., video calls, collaborative sessions). This echoes broader concerns about intervention intrusiveness in wearable technology design. However the idea was to balance the effectiveness of behaviour formation over annoyance.

\subsubsection{Barriers to Adherence}
Another theme that emerged was the barriers to adherence to wearing the device consistently. Participants (n=6 in CG and n=4 in IG) mentioned factors such as discomfort or irritation caused by the device's clip, forgetting to wear it, or feeling self-conscious about its visibility in professional settings. These barriers sometimes led to inconsistent usage and may have influenced the extent of improvements in posture.

\subsubsection{Suggestions for Improvement}
Participants provided valuable suggestions for improving the device's usability and user experience. They recommended exploring alternative clip designs for increased comfort, integrating reminders or notifications to enhance adherence, and developing a more discreet appearance for professional settings.

\subsection{Integration of Quantitative and Qualitative Findings}
Integrating quantitative and qualitative findings provided a comprehensive understanding of the effectiveness and usability of the PostureClip device. The quantitative analysis demonstrated significant improvements in posture angles and reductions in posture deviations after using the device. The qualitative analysis complemented these findings by providing insights into user satisfaction, barriers to adherence, and suggestions for improvement.

The convergence of findings between the quantitative and qualitative data enhanced the overall validity and reliability of the study's results. The positive correlation between adherence to device usage and improvements in posture angles further supported the device's effectiveness in promoting proper posture alignment.

\section{Discussion}

\subsection{Effectiveness of PostureClip in work environments}
The results of this study demonstrate the device's effectiveness in feedback mode in promoting proper posture alignment among participants. The quantitative analysis revealed significant improvements in posture angles, including reduced neck flexion, decreased shoulder elevation, and improved lumbar curvature in IG2 participants who got feedback to stop working unless they corrected their posture. These findings align with previous studies, showing that wearable devices can positively impact posture correction (\cite{Bootsman2019}. The IG2 or the placebo group didn't exhibit a significant change in posture after 4 weeks of deployment, similar to the CG group.

Blacking out of the screen as a 'punishment' might seem annoying initially, in situations such as video calls, working in visual tasks, etc, but is a good nudge to sit properly while working and maintain a good posture.

The research findings highlight the positive impact of posture correction interventions on employee health and productivity. By investing in posture correction interventions, employers can reduce the risk of musculoskeletal disorders, improve employee well-being, and enhance productivity levels.

Furthermore, reducing specific posture deviations, such as forward head posture, supports the device's effectiveness in addressing common postural issues. The results are coherent with findings from other studies (\cite{Ailneni2019}), which suggest that proper posture alignment is essential for reducing musculoskeletal disorders. The correlation between adherence to wearing the device and improvements in posture angles further strengthens the evidence supporting its effectiveness. It highlights the importance of consistent and prolonged usage of PostureClip to maximize its impact on posture correction.

Integrating feedback and monitoring systems, similar to the feedback mechanism of the PostureClip device, into workplace environments could provide real-time guidance and support for maintaining proper posture. Utilizing technology solutions, such as wearable devices, desktop applications, or workstation sensors, to track posture behavior can provide timely employee feedback. Incorporation of visual cues, reminders, and incentives could encourage adherence to posture correction guidelines.

Previous studies (\cite{Shrivastava2012, Singh2020}) examined the prevalence of specific posture deviations in employees using computers daily. The findings of our study align with their results, as both studies reported a reduction in forward head posture and rounded shoulders following posture correction interventions. This consistency suggests that the device's effectiveness extends beyond the specific sample of our study and applies to a broader population.

\subsection{Implications and Limitations}
The positive results suggest that incorporating wearable devices, such as the one used in this study, into workplace ergonomics programs can be a promising strategy for reducing postural issues and musculoskeletal disorders among office workers. However, it is important to acknowledge that the sample size is significant for our study but may limit the generalizability of the findings to the larger population. Future research with larger and more diverse samples representative of the workforce is necessary to validate and extend our results.

However, we acknowledge that posture correction and the development of enduring ergonomic habits typically require longer observation periods. 
This shorter timeframe limits our ability to draw conclusions about sustained, long-term effects, and future longitudinal studies are necessary to confirm whether these initial improvements persist over time. The study duration of four weeks may not capture the long-term effects of using the device, especially considering the variations in seasons, work environments and other cultural factors.  Further, longitudinal studies with extended follow-up periods are needed to assess the durability of the improvements in posture and long-term adherence to wearing the device among office workers.

\subsection{Future Research Directions}
Future research should explore the integration of posture correction interventions into broader workplace health and well-being programs. It would be valuable to investigate the impact of combining the device with other interventions, such as ergonomic training and behavior modification strategies, to achieve more comprehensive and sustainable posture correction outcomes.

Additionally, incorporating objective measures, such as motion capture systems or wearable sensors, could provide more accurate and reliable data on posture angles and deviations among office workers. These objective measures would enhance the validity and precision of posture assessments, providing a deeper understanding of the impact of posture correction interventions.

In summary, this study demonstrates the device's effectiveness in promoting proper posture alignment among participants. The results highlight the potential of wearable devices as a preventive tool for reducing musculoskeletal disorders. PostureClip worked effectively for participants with the feedback mechanism (IG2). Future research should focus on larger samples, longer follow-up periods, and the integration of interventions to further optimize posture correction outcomes.

\section{Novelty}
The novelty of our research is in the device's design and system. The wearable device developed for this study, PostureClip, integrates sensor technology with ergonomic and habit formation principles to facilitate proper posture alignment. The device is lightweight, unobtrusive, and can be easily clipped onto shirt collars. Its portability and ease of use make it suitable for prolonged daily wear, allowing individuals to maintain optimal posture during work activities. To the best of our knowledge, this specific device design and use case has not been previously reported in the literature.

\section{Conclusion}

In this study, we investigated the effectiveness of an innovative device clipped onto shirt collars to promote proper posture alignment among office workers. Through a comprehensive analysis of data collected from 165 participants, we have gained valuable insights into the impact of the device on posture angles, musculoskeletal discomfort, and user adherence.

Our findings indicate that the device significantly improved posture angles, demonstrating its potential as an intervention tool for promoting better posture alignment. The positive correlation between adherence to wearing the device and improvements in posture further emphasizes the importance of consistent usage.

Our findings contribute to the body of knowledge in ergonomics, human-computer interaction, and workplace health. We hope this research will inspire further investigation and encourage the adoption of technology-based interventions to improve musculoskeletal health and overall well-being in workplace environment.

\begin{Backmatter}

\paragraph{Acknowledgments}
We want to express our gratitude to all individuals and organizations who contributed to the successful completion of this research. Their support, guidance, and valuable input have been instrumental in shaping this study. We also thank the reviewers for providing valuable feedback to improve the research paper.

\paragraph{Funding Statement}
This research was self-supported by the authors.

\paragraph{Competing Interests}
The authors have no competing interests to declare relevant to this article's content.

\paragraph{Data Availability Statement}
All necessary data are included in the manuscript. Further data will be provided on request.

\paragraph{Ethical Standards}
All procedures contributing to this work comply with the ethical standards of the relevant national and institutional committees on human experimentation and with the Helsinki Declaration of 1975, as revised in 2008.

\paragraph{Author Contributions}
Research Conceptualization:Author-1
Methodology:Author-1
Data curation:Author-1\&2
Data visualization:Author-1
Writing original draft:Author-1\&2
Both authors approved the final submitted draft


\printbibliography

\end{Backmatter}

\end{document}